\documentclass[twocolumn,twoside,slac_two]{revtex4}
\usepackage{graphicx}
\usepackage{fancyhdr}
\newcommand{\be}{\begin{equation}}
\newcommand{\ee}{\end{equation}}

\begin{document}

\title{Diffuse Galactic Gamma Rays at Intermediate and High Latitudes, Constraints on ISM Properties}

%

\author{Maryam Tavakoli}
\affiliation{SISSA, Via Bonomea, 265, 34136 Trieste, Italy}
\affiliation{INFN, Sezione di Trieste, Via Bonomea 265, 34136 Trieste, Italy}
\author{Ilias Cholis} 
\affiliation{SISSA, Via Bonomea, 265, 34136 Trieste, Italy}
\affiliation{INFN, Sezione di Trieste, Via Bonomea 265, 34136 Trieste, Italy}
\author{Carmelo Evoli}
\affiliation{SISSA, Via Bonomea, 265, 34136 Trieste, Italy}
\affiliation{National Astronomical Observatories, Chinese Academy of Sciences, 20A Datun Road, Beijing 100012, P.R. China}
\author{Piero Ullio}
\affiliation{SISSA, Via Bonomea, 265, 34136 Trieste, Italy}
\affiliation{INFN, Sezione di Trieste, Via Bonomea 265, 34136 Trieste, Italy}

\begin{abstract}
The  spectral data on the diffuse Galactic $\gamma$-rays, at medium and high latitudes 
($\mid b \mid > 10^{\circ}$) and energies of 1-100 GeV, recently published by the 
\textit{Fermi} Collaboration are used to produce a novel study on the $\gamma$-ray
emissivity in the Galaxy. We focus on analyzing the properties of propagation of cosmic rays 
(CRs), using the publicly available DRAGON code. We critically address some of the models 
for the interstellar HI and H2 gas distributions commonly used in the literature, as well as test 
a variety of propagation models.
Each model assumes a distinct global profile for the diffusion and the re-acceleration of CRs.
Fitting propagation parameters to well measured local CRs such as, the B/C ratio, 
$p$, $\bar{p}$ and $e^{\pm}$ fluxes, we evaluate the $\gamma$-ray spectra at medium and 
high latitudes in order to place further constraints on these propagation models.

\end{abstract}

\maketitle

\thispagestyle{fancy}

\section{Introduction}
Interactions of CRs with the interstellar medium (ISM) are a copious source of $\gamma$-rays through $\pi^0$, inverse Compton and Bremsstrahlung emissions.
The study of diffuse $\gamma$-rays at intermediate and high latitudes ($|b| > 10^{\circ}$) \cite{Abdo:2010nz} is a promising tool to probe and constrain the propagation of CRs in the Galaxy as well as ISM properties, with the CR propagation described by:
\begin{eqnarray} 
\frac{\partial \psi}{\partial t} &=& q(\vec{r},p,t)+\vec{\nabla}.(D_{xx}\vec{\nabla}\psi) \nonumber \\ 
&+&\frac{\partial}{\partial p}\Big[p^2D_{pp}\frac{\partial}{\partial p}(\frac{\psi}{p^2})\Big] -\frac{\partial}{\partial p}(\dot{p}\psi)-\vec{\nabla}.(\vec{V}\psi)\nonumber \\
&+&\frac{\partial}{\partial p}\Big[\frac{p}{3}(\vec{\nabla}.\vec{V})\psi\Big] - \frac{\psi}{\tau_{frag}}-\frac{\psi}{\tau_{decay}}\,,
\label{eq:CR_trans}
\end{eqnarray}
where $\psi(\vec{r},p,t)$ is the CR density, $q(\vec{r},p,t)$ is the source term 
including components of primary origin, as well as CRs from spallation and decay processes from heavier elements. $D_{xx}(\vec{r})$ is the spatial diffusion,  $D_{pp}(\vec{r})$ describes the diffusion in momentum space due to re-acceleration,  $\dot{p}$ is the momentum 
loss rate due to interactions with ISM, the Galactic magnetic field 
or the interstellar radiation field (ISRF), $\vec{V}$ is the convection velocity due to Galactic winds, while 
$\tau_{frag}$ and $\tau_{decay}$ are the timescales for, respectively, fragmentation loss 
and radioactive decay.

For our simulations we use the DRAGON code~\cite{Evoli:2008dv} that numerically solves Eq.~(\ref{eq:CR_trans}) in the steady state approximation, 
in a 3D grid; 2 spatial dimensions for Galactocentric radial distance and height from the Galactic plane, and 1 for the momentum $p$. 
\section{Assumptions}
\subsection{Primary Sources} 
CR primary sources up to energies of $\sim$~100TeV, are supernova remnants (SNRs). 
For each nucleus $i$ the source term describing the injection of CRs in the ISM is 
given as a function of rigidity R by
\be 
q_i(r,z,E)=q_{0,i}f_s(r,z)(\frac{R(E)}{R_0})^{-\gamma^i}\,,
\ee
where $q_{0,i}$ is the normalization for each nucleus, $f_s(r,z)$ traces the distribution of 
SNRs as modeled in \cite{Ferriere:2001rg} on the basis of pulsar and progenitor star surveys 
\cite{Evoli:2007iy}. 
Electrons and positrons accelerated between a pulsar and the termination shock
of the wind nebula may contribute to the high energy $e^{\pm}$ spectrum, and then to the $\gamma$-ray flux, via inverse Compton scattering.
Following the parametrization of \cite{Malyshev:2009tw}, the source term due to a distribution of pulsars can be described by a power-law with an exponential cut-off given by
\be
Q_{p}(r,z,t,E) = J_{0} E^{-n} e^{-E/M} f_{p}(r,z)\,, 
\label{eq:PulsarSource}
\ee
where $J_0$ depends on the averaged birth rate of pulsars distribution and the average portion of pulsar initial rotational energy injected in the ISM as $e^{\pm}$, $f_p(r,z)$ describes the spatial distribution of young and middle age pulsars modeled by \cite{FaucherGiguere:2005ny}, while $n$ and $M$ are, respectively, the injection index and statistical cut-off, for the pulsar distribution.
\subsection{Magnetic Field and Diffusion} 
The large scale Galactic magnetic field is generally assumed to be a bi-symmetrical spiral with a small pitch angle \cite{Jansson:2009ip}. 
Here we assume that the regular magnetic field is purely azimuthal, $ \vec{B_0}=B_0 \hat{\phi}$, and has the form
\begin{equation} 
B_0=3\exp{(-\frac{r-r_{\odot}}{11(\textrm{kpc})})}\exp{(-\frac{|z|}{2(\textrm{kpc})})} (\mu {\rm G})\,, 
\end{equation} 
based on the analysis of WMAP synchrotron intensity and polarization data in \cite {MivilleDeschenes:2008hn}.

Diffusion is a result of CR scattering on randomly moving magneto-hydro-dynamical (MHD) waves. A correlation to the Galactic magnetic field and its discontinuities is expected, with a larger diffusion coefficient where the magnetic field is weaker. We assume, on the basis of \cite{Evoli:2008dv}, an isotropic diffusion coefficient of the form 
\be 
D(r,z,R)=D_0 \beta ^{\eta}(\frac{R}{R_0})^{\delta}\exp{(\frac{r-r_{\odot}}{r_d})}\exp{(\frac{|z|}{z_d})}\,,
\label{eq:DiffCoef}
\ee
where $R_0=3~GV$ is the reference rigidity, $\delta$ is the diffusion spectral index depending on mechanism that builds up the turbulence in the ISM, $r_d$ and $z_d$ are, respectively, the radial and vertical scales defining the diffusion profiles in the Galaxy. The dependence of diffusion on the particle velocity, $\beta=v_p/c$, is naturally expected to be linear ($\eta=1$), 
however the analysis by \cite{Ptuskin:2005ax} shows an increase in diffusion at low energies. 
To represent such a behavior, the parameter $\eta$ has been introduced by \cite{DiBernardo:2010is}.\\
\subsection{Interstellar Gas:}  
The interstellar gas (ISG) is composed of hydrogen, helium and small contributions from heavier elements, 
with hydrogen observed in atomic (HI), molecular (H2) and ionized (HII) states. For the distribution of HI gas we use as a reference the model developed by \cite{Nakanishi:2003eb}. 
 
\begin{figure}
\begin{center}
\includegraphics[width=80mm]{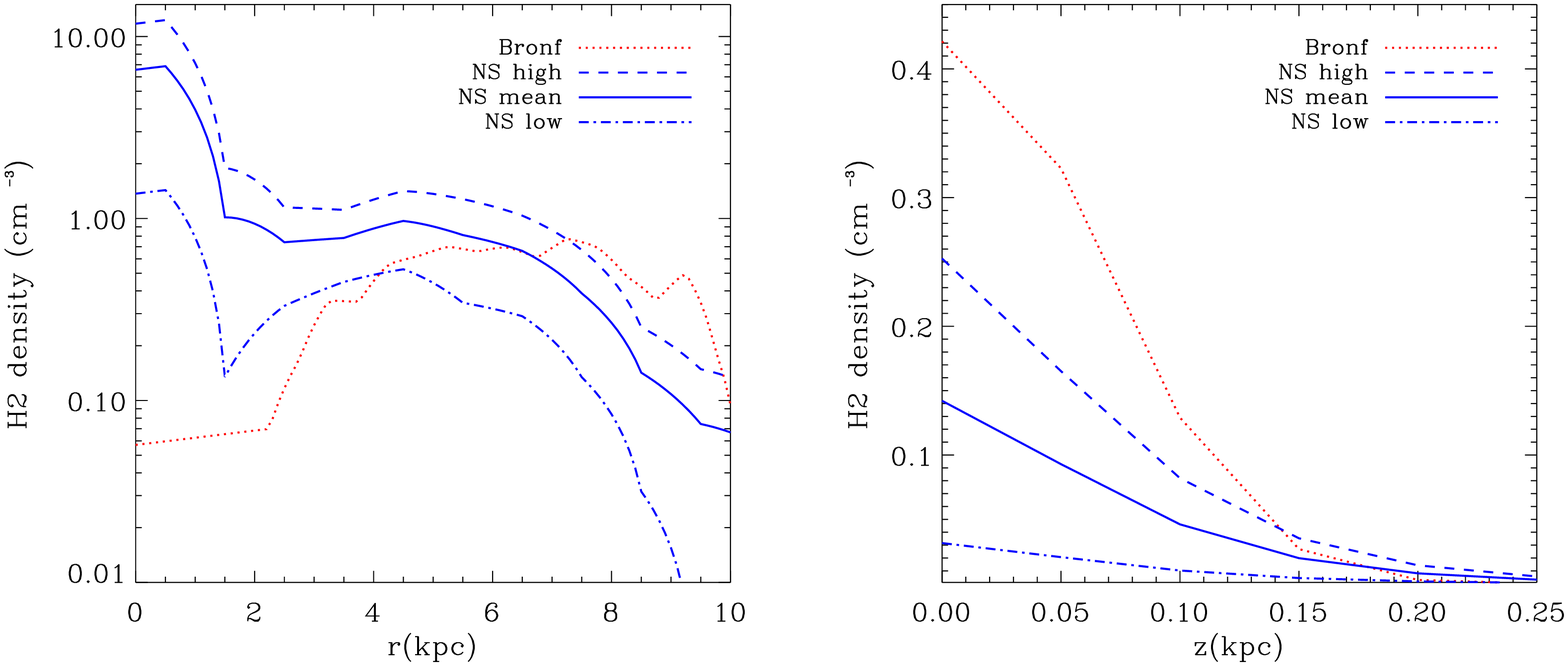}
\end{center}
\caption{  
Large scale molecular hydrogen distribution in the Galaxy vs $r$ for $z=0$(\emph{left}); vs $z$ for $r=r_{\odot}$ (\emph{right}). \emph{dashed}, \emph{solid} and \emph{dashed-dotted} lines for our reference model \cite{Nakanishi:2006zf} with, respectively, high, mean and low values of H2 mid plane density, \emph{dotted} lines for the model provided by \cite{1988ApJ...324..248B}.}
\label{fig:Gasprofile}
\end{figure}

Molecular hydrogen can exist only in dark cool clouds where it is protected against the ionizing stellar ultraviolet radiation. 
It can be traced with the $\lambda$ = 2.6 mm (J = 1 $\to$ 0) emission line of CO, since collisions between the CO and H2 molecules in the clouds are responsible for the excitation of CO. The CO to H2 conversion factor, $X_{CO}$ which relates the H2 column density, $N_{H2}$ , to the velocity-integrated intensity of the CO line, has considerable uncertainties. 
For the H2 distribution we use for our reference model the map provided by \cite{Nakanishi:2006zf},
assuming the conversion factor to vary exponentially with Galactocentric radius,
\begin{equation} 
\textrm{X}_{\textrm{CO}}[\textrm{H}_2 \textrm{cm}^{-2} \textrm{K}^{-1} \textrm{km}^{-1} \textrm{s}]=1.4\exp(\frac{R}{11(\textrm{kpc})})\,, 
\end{equation}
however \cite{1988ApJ...324..248B} is also an older widely used model in the literature. The radial and vertical profiles of H2 distribution models are shown in Fig.~\ref{fig:Gasprofile}. 
\newcommand{\imsize}{1\columnwidth}
Ionized hydrogen occurs in the vicinity of young O and B stars, with the ultraviolet radiation from these stars 
ionizing the ISM. HII regions have a similar distribution to the molecular hydrogen, but mass-wise their 
contribution is negligible. 
\subsection{Methodology}
\label{sec:astro_analysis}
\begin{figure}
\begin{center}
\includegraphics[width=34.2mm]{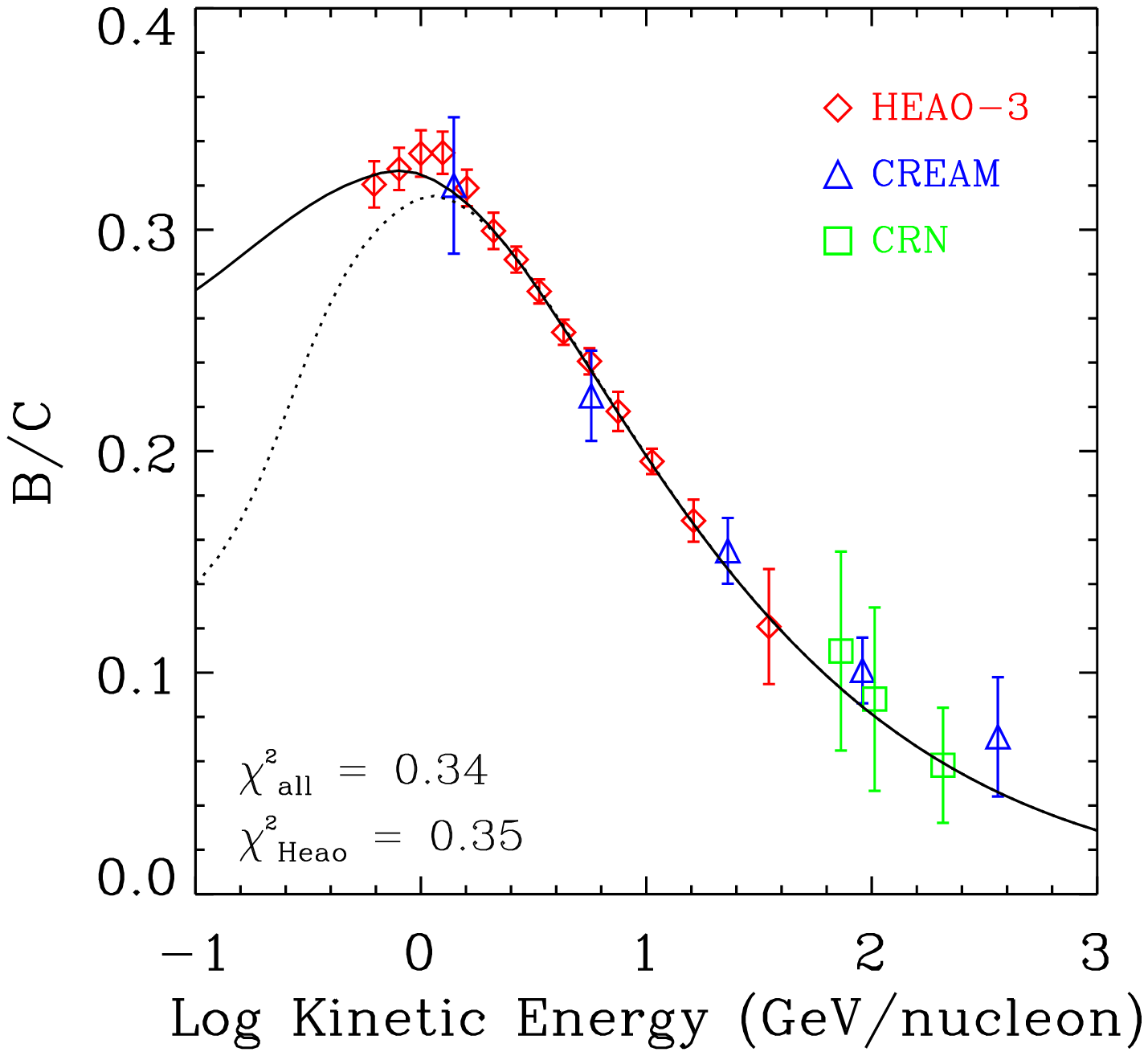} 
\includegraphics[width=45.8mm]{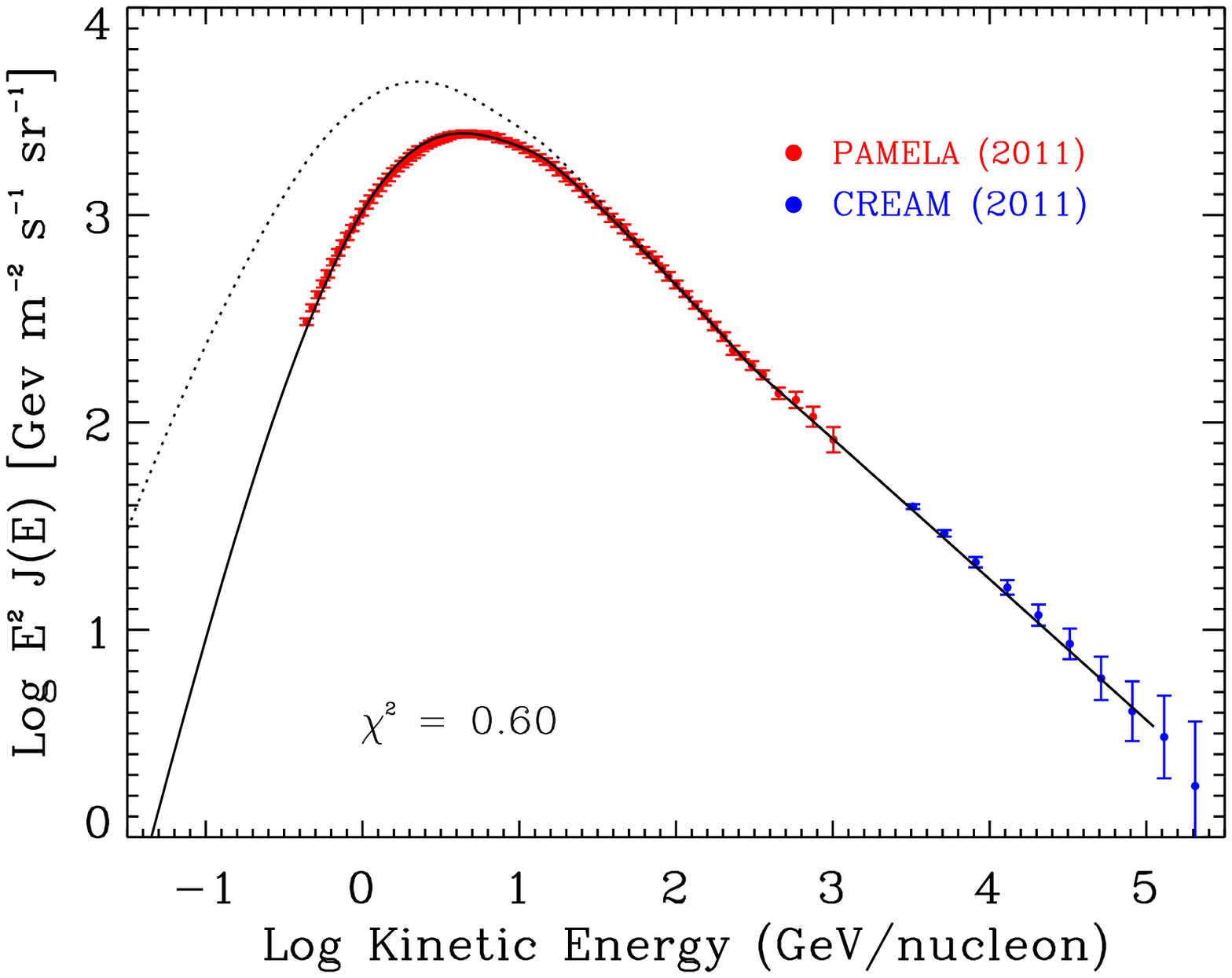} \\
\includegraphics[width=34.2mm]{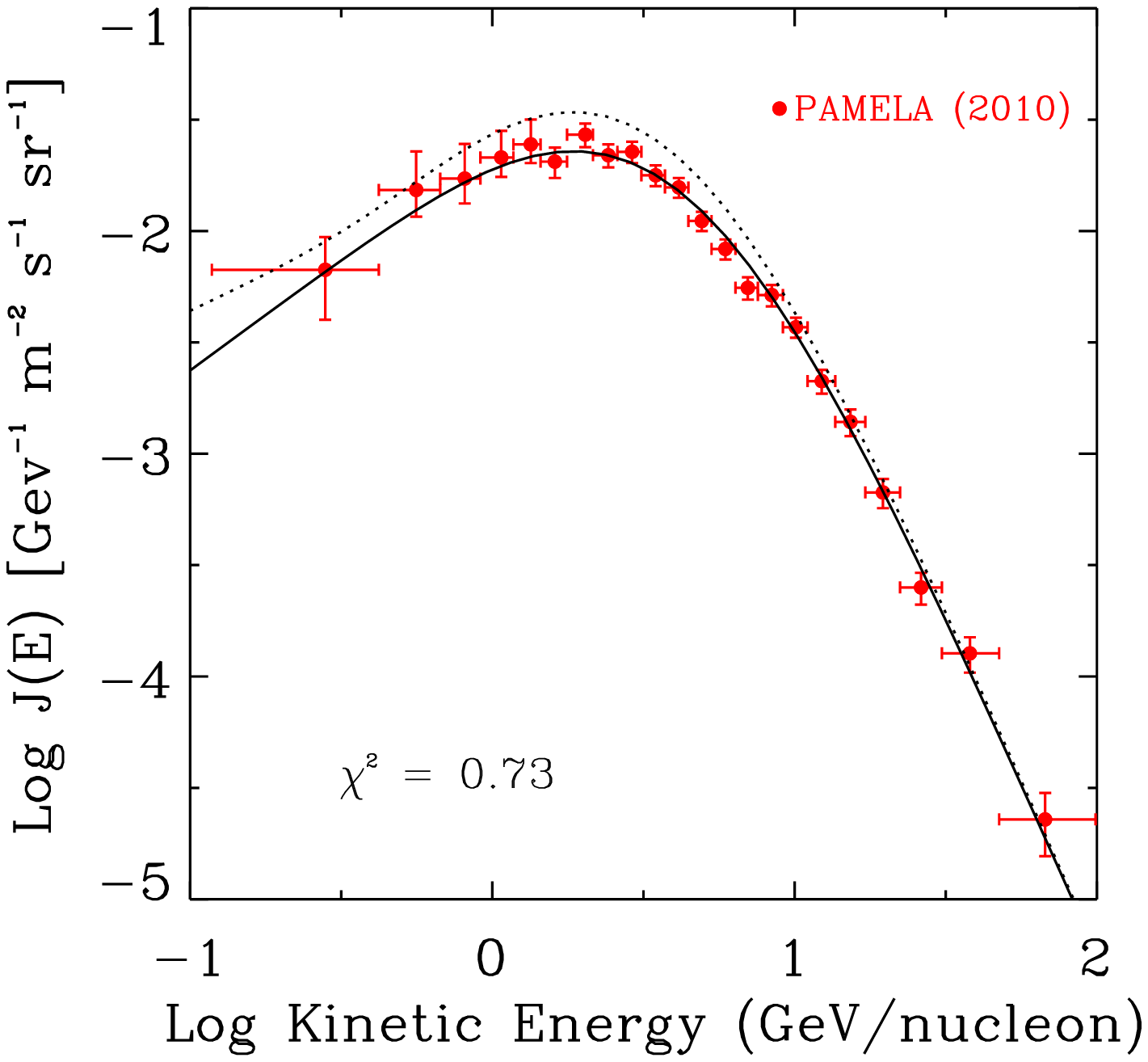}
\includegraphics[width=45.8mm]{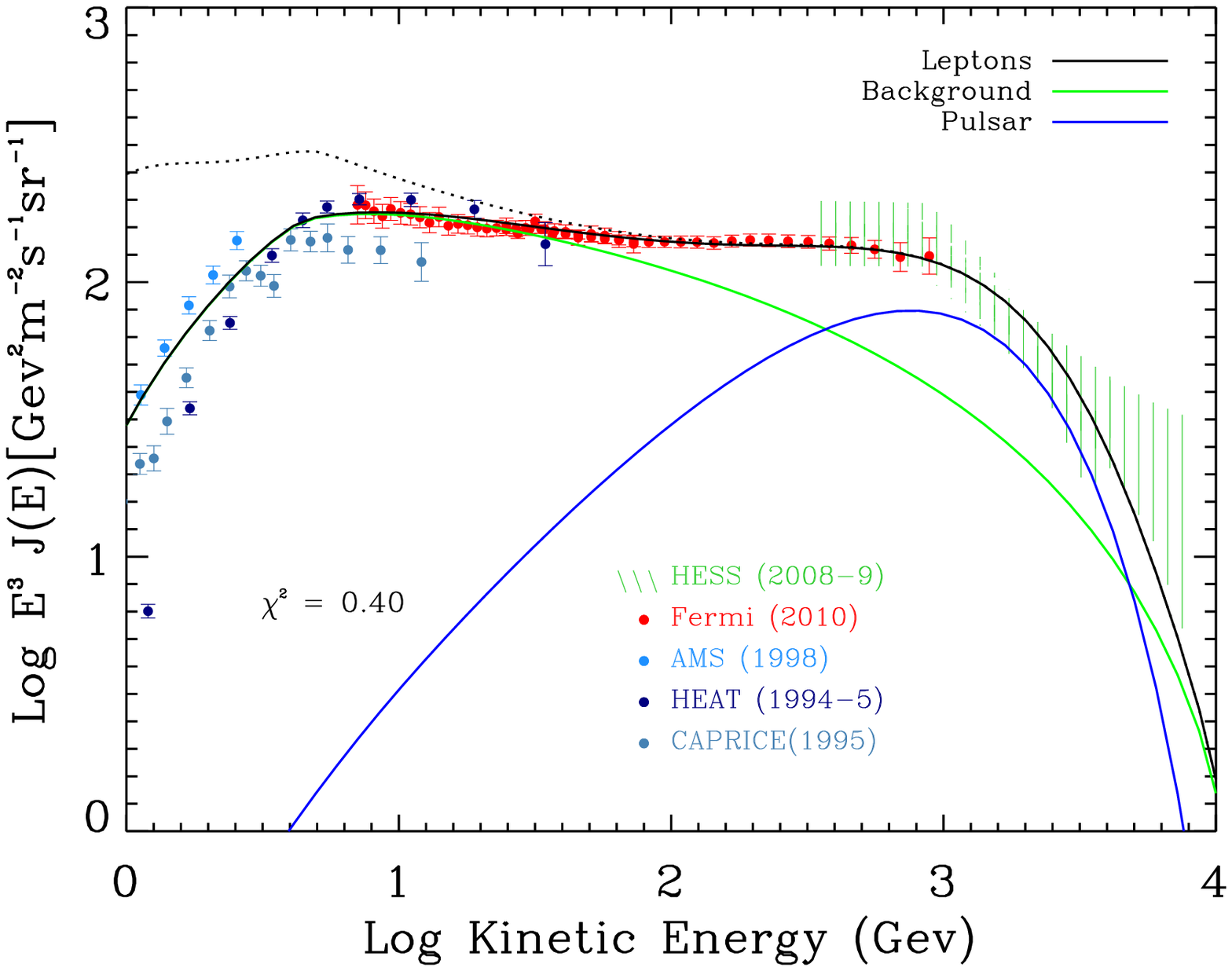}
\end{center}
\caption{
\emph{Upper left}:  B/C spectrum to fit the diffusion parameters in  Eq.~(\ref{eq:DiffCoef}),
\emph{Upper right}: the proton flux to fit $\gamma^{p}_{i}$ ,
\emph{Lower left}: the predicted antiproton spectrum,  
\emph{Lower right}: the $e^{+}+e^{-}$ flux to fit $\gamma^e$ and pulsar parameters in Eq.~(\ref{eq:PulsarSource}).}
\label{fig:RefModelCRs}
\end{figure}

The propagation parameters are determined upon fitting CR spectra. We consider a range of values for  $\delta$, $z_d$ and $r_d$ in Eq.~(\ref{eq:DiffCoef}). For each set of these values we derive ($D_0, v_A, \eta$) by minimizing the $\chi^2$ for Boron over Carbon $B/C$ data (Fig.~\ref{fig:RefModelCRs} (upper left)).

The injection spectrum of protons described by three spectral indices, $\gamma^{p}_{i}$ (see Table~\ref{tab:Param}) is fitted to the  \textit{PAMELA} \cite{Adriani:2011cu} and CREAM data \cite{Yoon:2011zz} (Fig.~\ref{fig:RefModelCRs} (upper right)).
The predicted antiproton spectrum is consistent with local flux (Fig.~\ref{fig:RefModelCRs} (lower left)).
Helium spectrum is also checked for consistency with the most recent data from  \textit{PAMELA} \cite{Adriani:2011cu}.

The $e^{\pm}$ fluxes below  $E \sim$ 30 GeV is dominated by primary electrons accelerated by supernovae and by secondary electrons (and positrons) produced in inelastic collisions of CR nuclei with the ISM. The spectrum of the secondary $e^{\pm}$ is related to the CR nuclei spectrum, while the primary electron spectrum can be described by a single power-law $\gamma^e$ above 5 GeV; which we fit from the low energy $e^{+} + e^{-}$ spectrum measured by \textit{Fermi} (for a more detailed discussion see \cite{Cholis:2011un}). Pulsars within $\sim 3$kpc may also contribute to the $e^{+} + e^{-}$ spectrum up to $O(0.1)$ at $E \approx 50$~GeV
and up to $O(1)$ at $E \approx 500$~GeV \cite{Malyshev:2009tw}. 
Assuming pulsars contribute maximally, we find the properties of pulsars $J_0$, $n$, and $M$ in Eq.~(\ref{eq:PulsarSource}) upon fitting the \textit{Fermi} data as shown in Fig.~\ref{fig:RefModelCRs} (lower right). The positron fraction and electron spectra measured by \textit{PAMELA} are checked for consistency as well. 
\section{Results}
\label{sec:astro_results}
\begin{figure}
\begin{center}
\includegraphics[width=40mm]{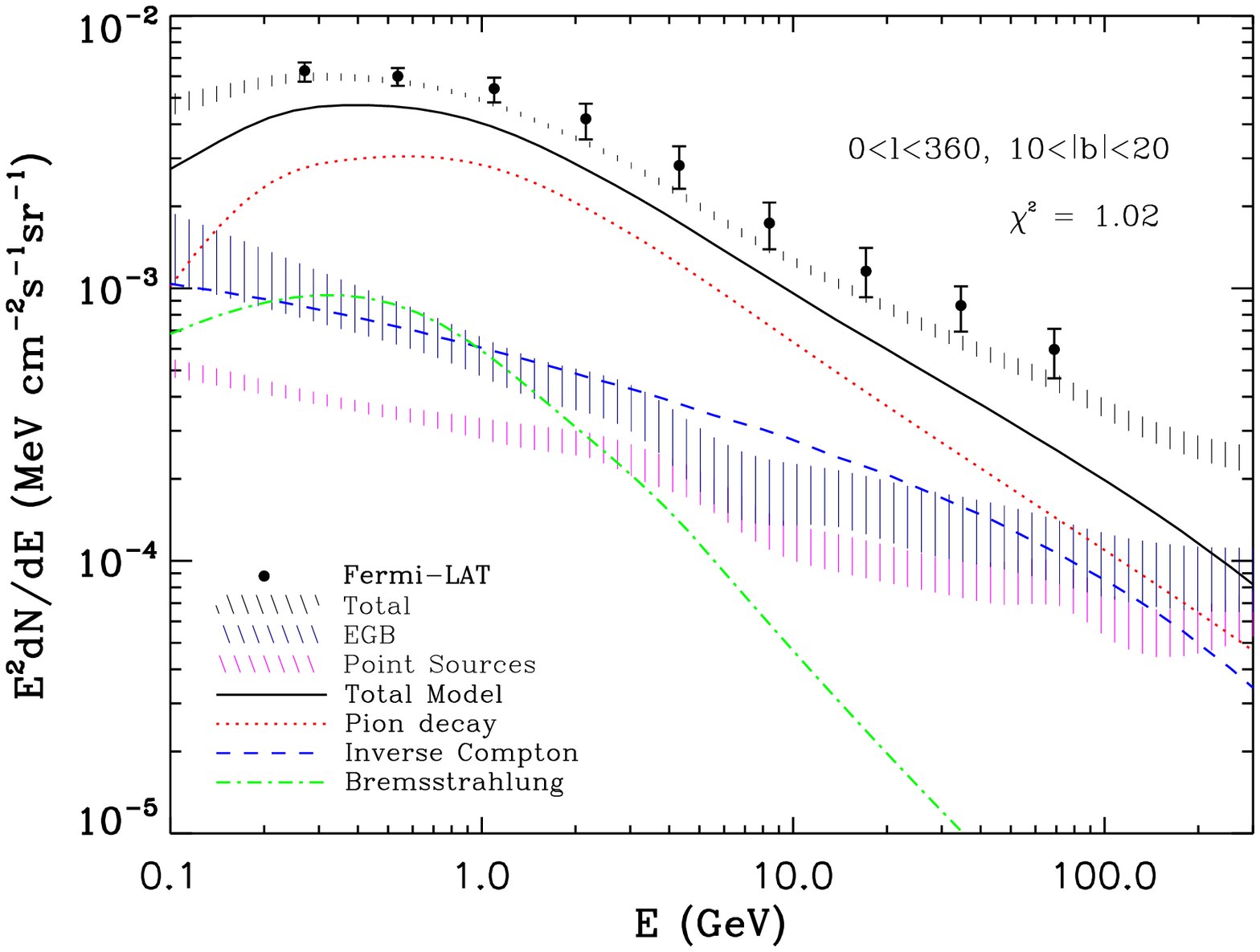}
\includegraphics[width=40mm]{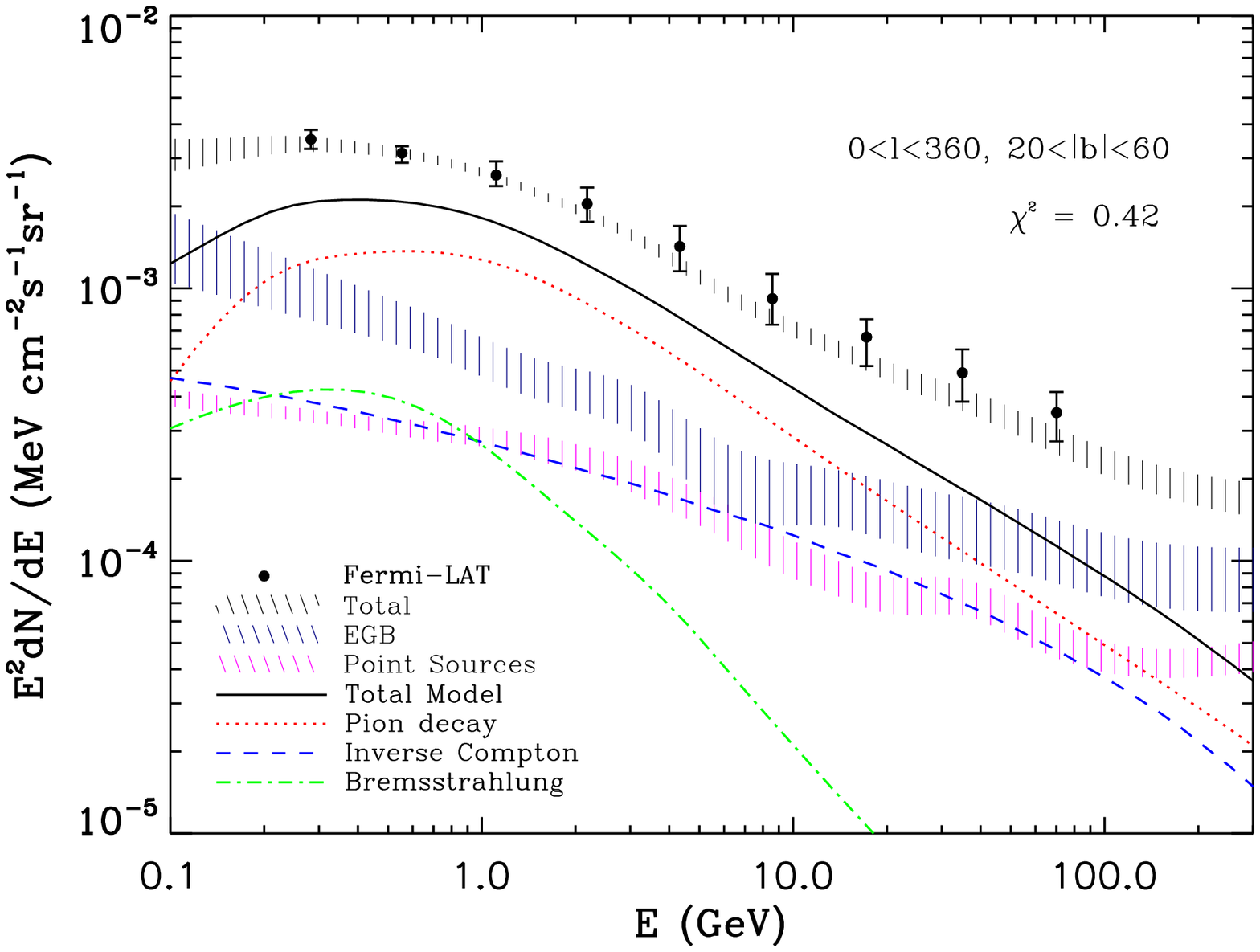} \\
\includegraphics[width=40mm]{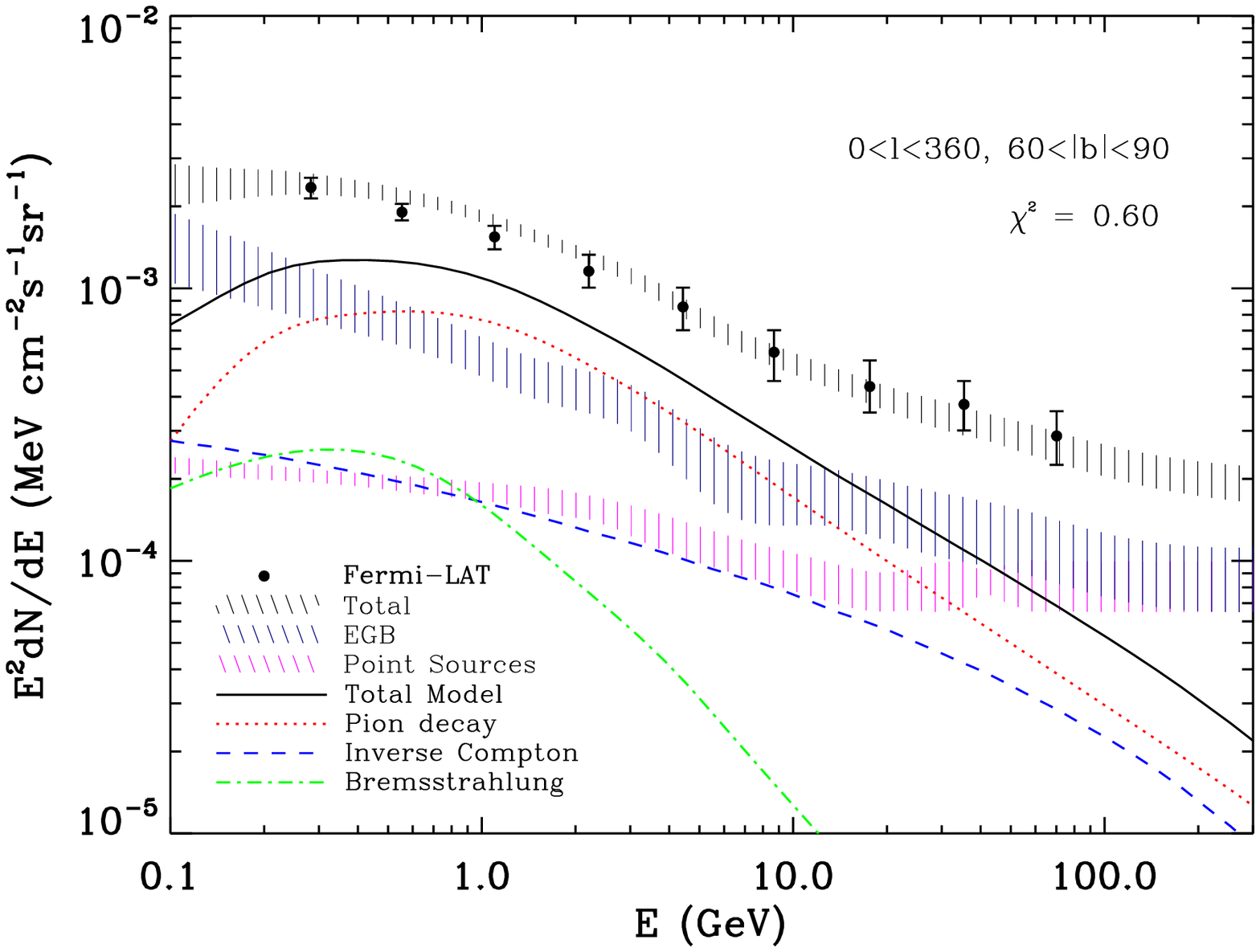}
\end{center}
\caption{Reference model, predictions for the $\gamma$-ray spectrum for $0^{\circ} < l < 360^{\circ}$, 
 \emph{Upper left}: $10^{\circ} < |b| < 20^{\circ}$,
\emph{upper right}: $20^{\circ} < |b| < 60^{\circ}$,
\emph{lower}: $60^{\circ} < |b| < 90^{\circ}$.}
\label{fig:RefAstroModelGammas}
\label{fig:gamma}
\end{figure}
Our reference model with $\delta = 0.5$, $z_{d} =4$~kpc and $r_{d} =20$~kpc called "KRA4-20" provides a good combined fit of local CRs (Fig.~\ref{fig:RefModelCRs}) and diffuse $\gamma$-ray spectra at intermediate and high latitudes (Fig.~\ref{fig:gamma}).

\begin{figure}
\begin{center}
\includegraphics[width=40mm]{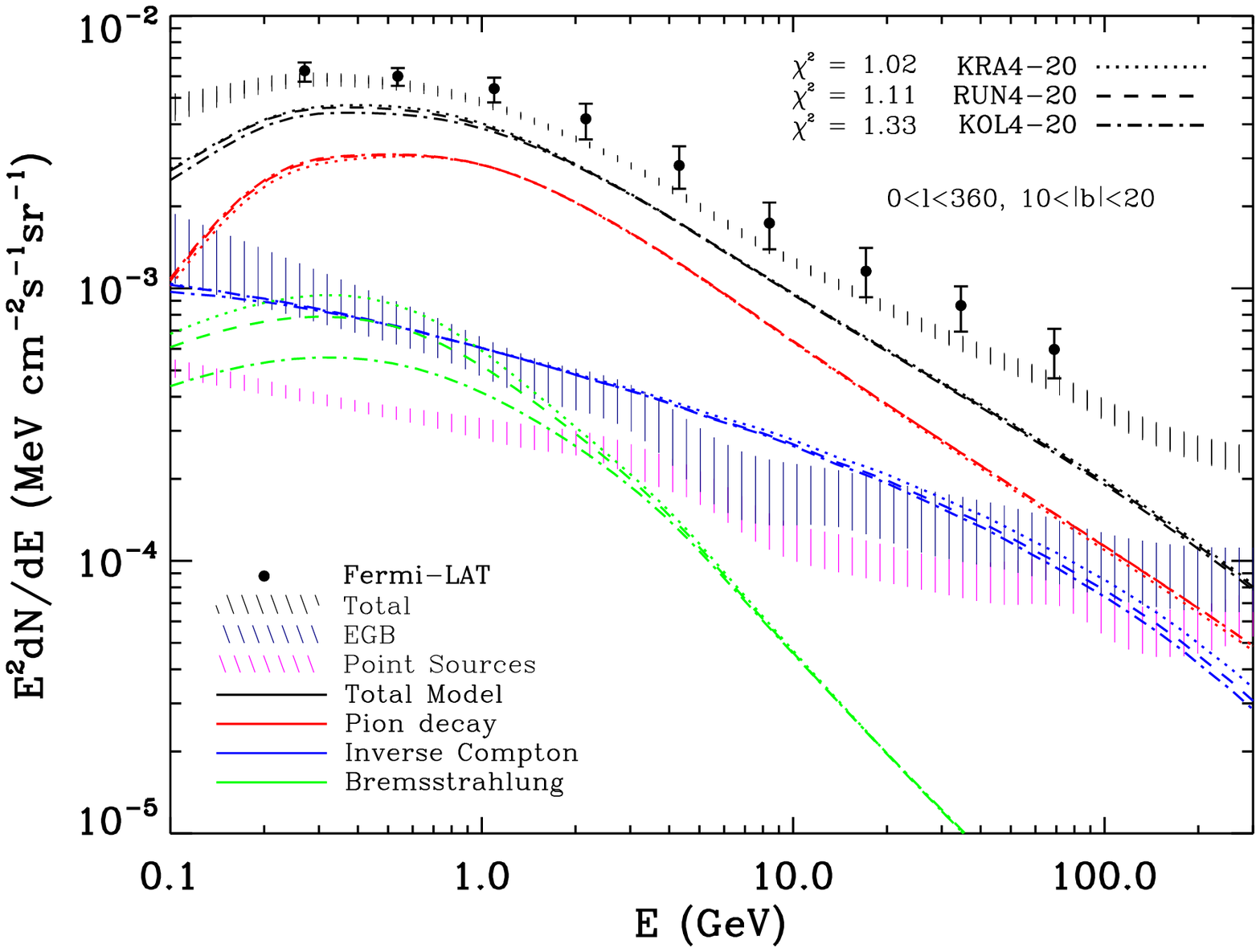}
\includegraphics[width=40mm]{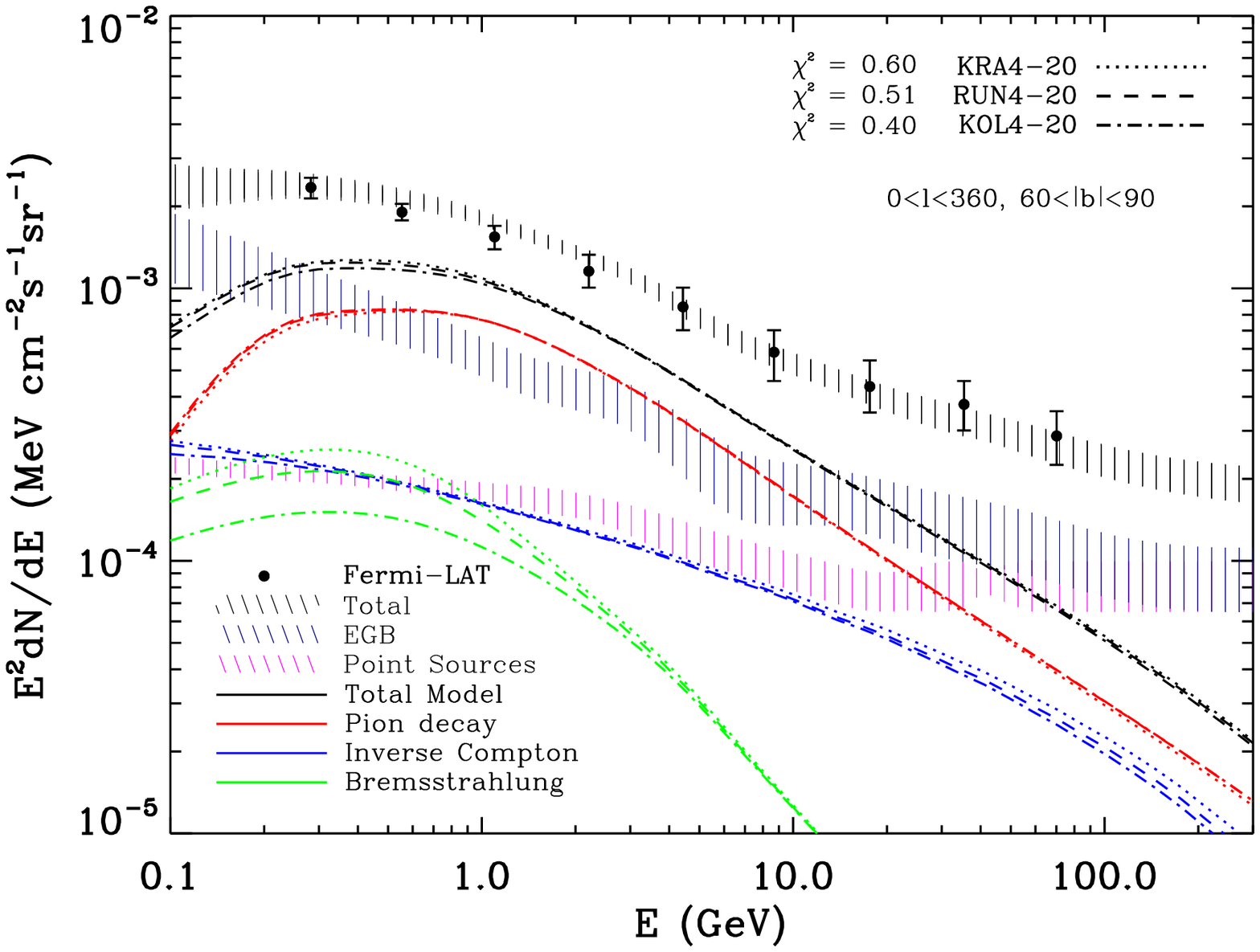}
\end{center}
\caption{Predictions for different values of the diffusion index $\delta$. \emph{dotted lines}: $\delta = 0.5$, \emph{dashed lines}: $\delta = 0.4$, \emph{dashed-dotted lines}: $\delta = 0.33$. For all $z_d=4$~kpc and $r_d=20$~kpc.}
\label{fig:DiffIndexVary}
\end{figure}
In Fig.~\ref{fig:DiffIndexVary} we show the $\gamma$-ray spectra for different spectral indices; $\delta=0.5$ ("KRA4-20"), $\delta=0.4$ ("RUN4-20") and $\delta=0.33$ ("KOL4-20"). The $\pi^0$ component depends on the propagated proton spectrum. Lower values of the $\delta$ makes the propagated spectrum harder, resulting in the need for a softer injection index $\gamma^{p}_{i}$ in order to fit the data as shown in Table~\ref{tab:Param}. 
The differences in $\gamma^{p}_{i}+\delta$ causes small differences in the $\pi^0$ fluxes. 
Electrons propagation, unlike protons, mainly depends on energy loss time-scale and thus is not affected much by varying the diffusion index. 
Since we normalize our diffusion coefficient at 3 GV (see Table~\ref{tab:Param}), 
for larger $\delta$ the higher energy $e^{\pm}$ diffuse faster, reaching the 
higher latitudes faster giving a slightly harder inverse Compton spectrum. 
Differences in low energy Bremsstrahlung emissions come from different Alfv$\acute{\textrm{e}}$n velocities (see Table~\ref{tab:Param}), with the greater re-acceleration depleting the low energy spectrum.
\begin{figure}
\begin{center}
\includegraphics[width=40mm]{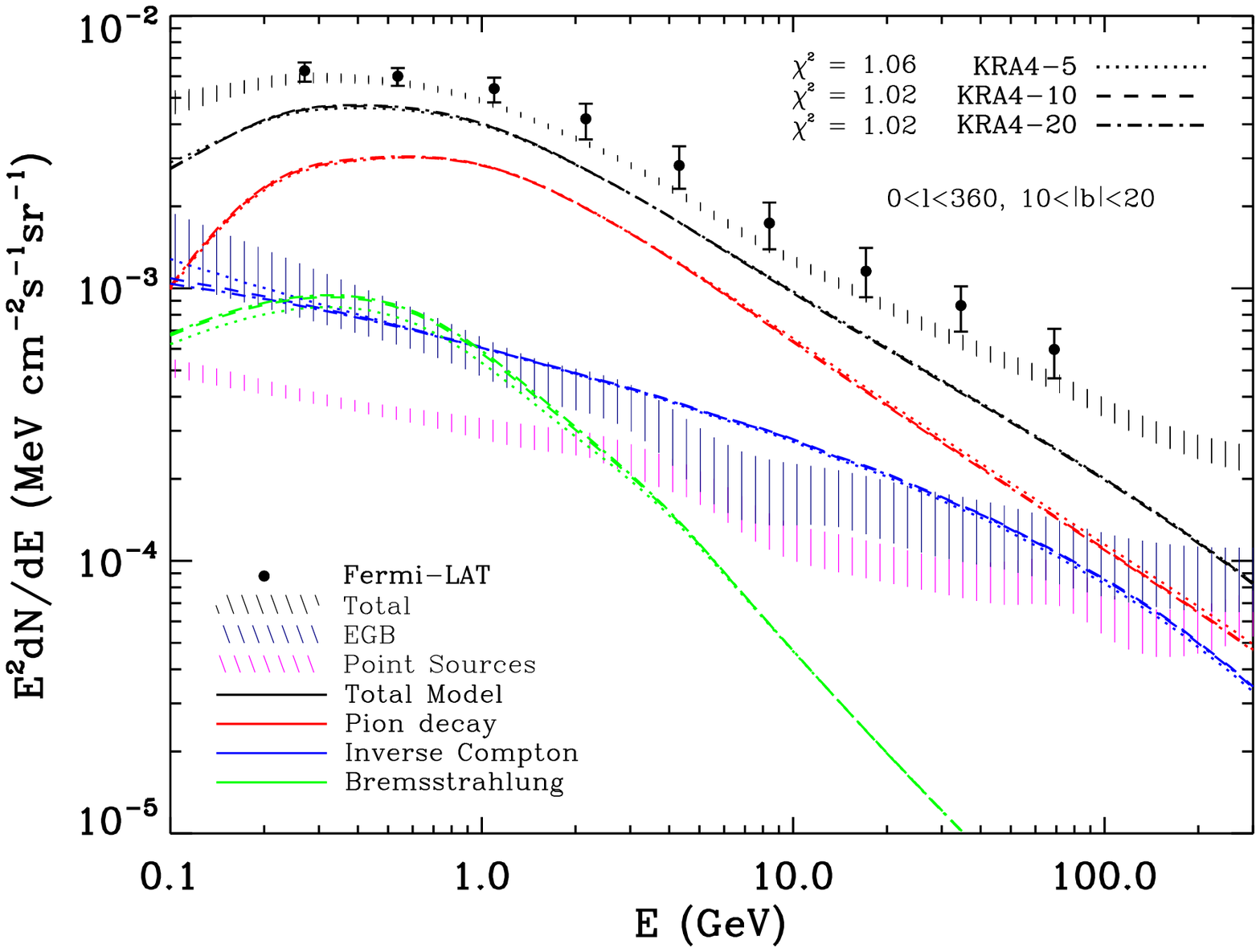}
\includegraphics[width=40mm]{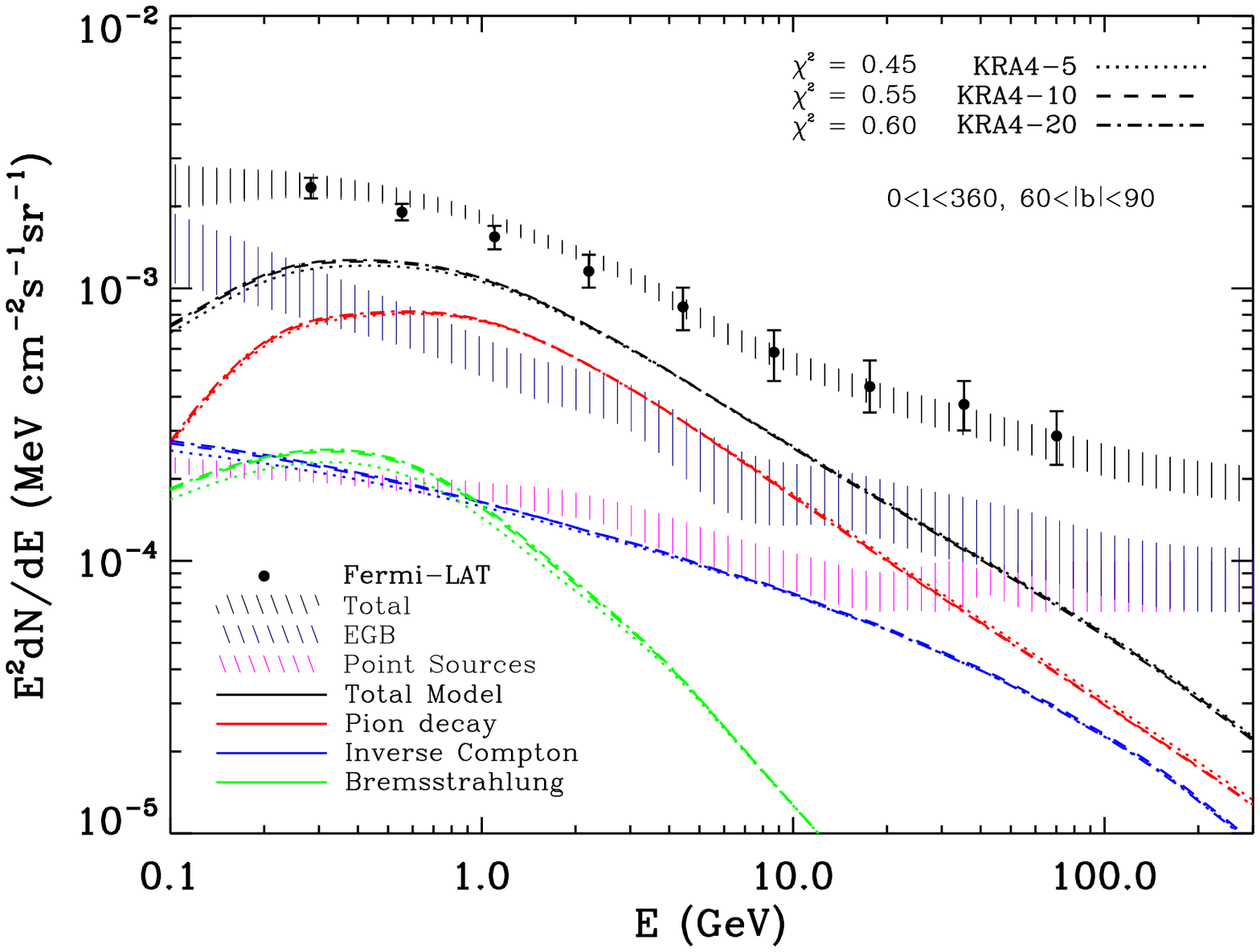}
\end{center}
\caption{Predictions varying the diffusion radial scale $r_{d}$. \emph{dotted lines}: $r_{d} = 5$~kpc, \emph{dashed lines}: $r_{d} = 10$~kpc,
 \emph{dashed-dotted lines}: $r_{d} = 20$~kpc. For all $\delta=0.5$ and $z_d=4$~kpc.}
 \label{fig:rdVary}
\end{figure}
The decreasing of $r_d$ from $20~kpc$ ("KRA4-20") to $5~kpc$ ("KRA4-5") makes the diffusion coefficient smaller towards the Galactic center, which forces CRs produced closer to the Galactic center to spend greater time there. Since we refit the $D_0$ for each propagation model (see Table ~\ref{tab:Param}) the net effect in the fluxes is negligible (Fig.~\ref{fig:rdVary}).   
\begin{figure}
\begin{center}
\includegraphics[width=40mm]{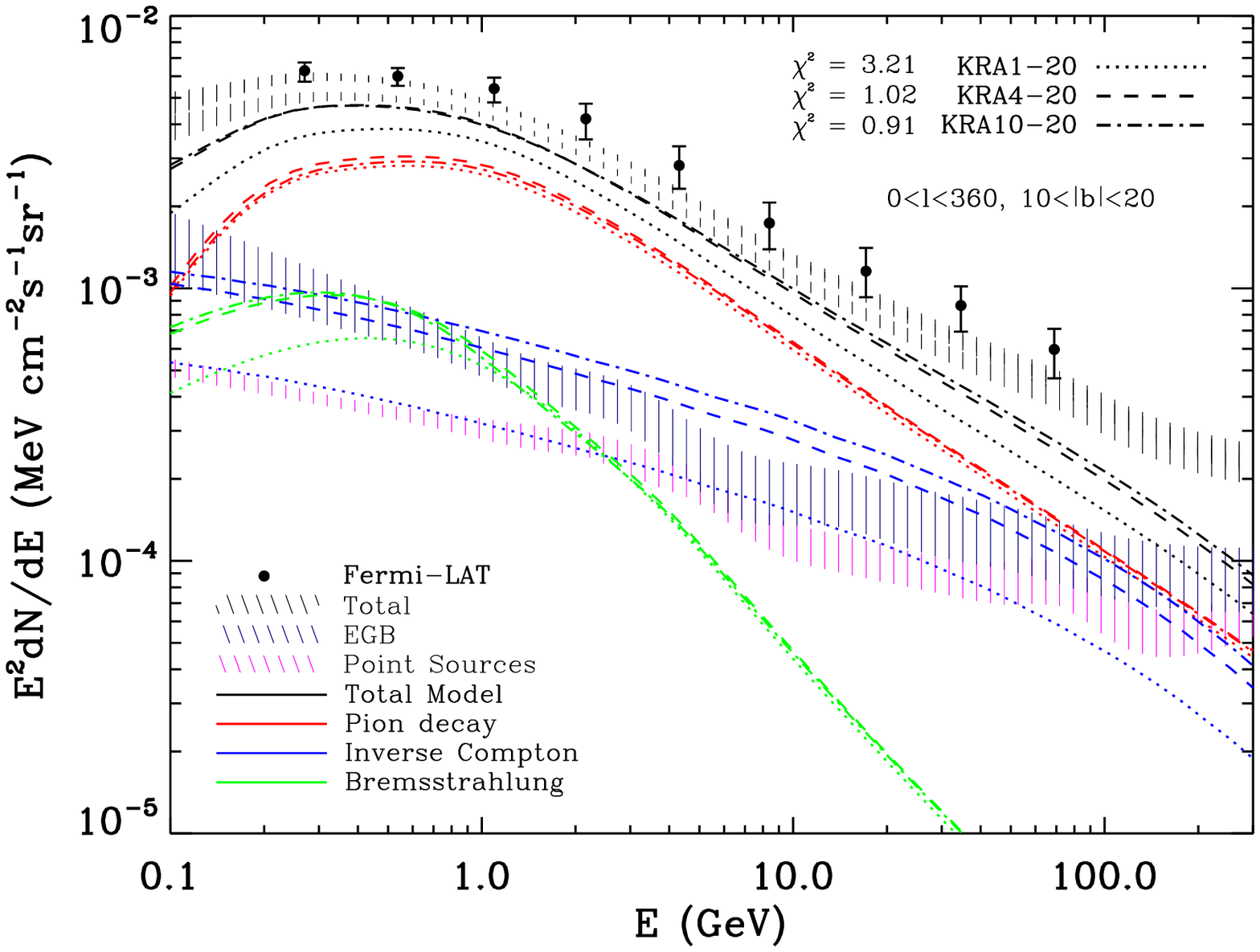}
\includegraphics[width=40mm]{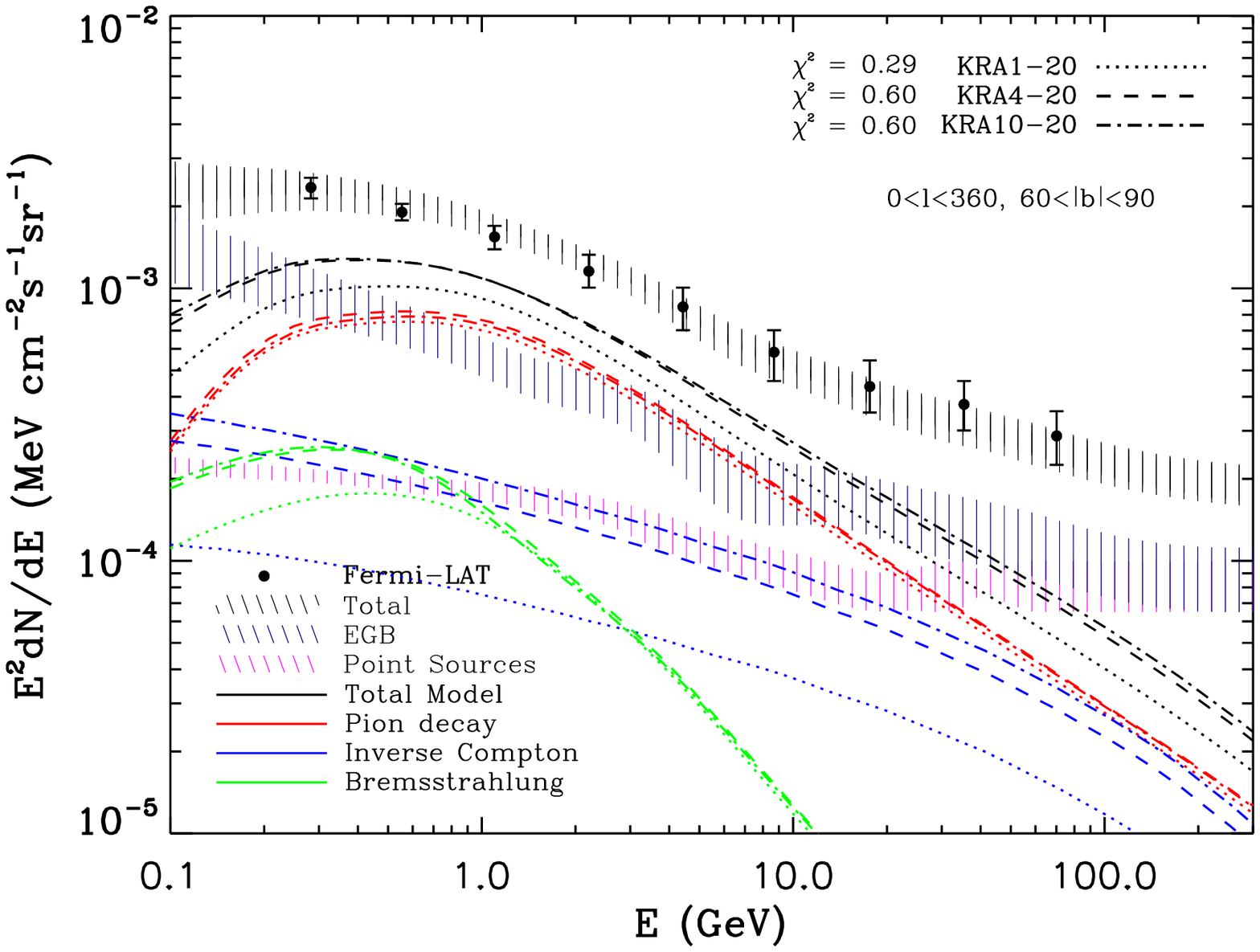}
\end{center}
\caption{Predictions varying the diffusion scale height $z_{d}$. \emph{dotted lines}: $z_{d} = 1$~kpc, \emph{dashed lines}: $z_{d} = 4$~kpc,
 \emph{dashed-dotted lines}: $z_{d} = 10$~kpc. For all $\delta=0.5$ and $r_d=20$~kpc.}
 \label{fig:zdVary}
\end{figure}
In Fig.~\ref{fig:zdVary} we show the effect of varying the diffusion scale height $z_{d}$ from $1~kpc$ ("KRA1-20") to $10~kpc$ ("KRA10-20"). Since $\pi^0$ and Bremsstrahlung emissions are morphologically correlated to the gas distribution which is concentrated close to the galactic disk, they do not change much by changing the size of diffusion zone. The inverse Compton spectrum is mainly affected by the actual distribution of 
electrons being confined within thinner (thicker) diffusion zones resulting
in lower (higher) total inverse Compton flux, since even infrared and optical 
target photons have a much thicker distribution profile than the ISM gas.
\begin{figure}
\begin{center}
\includegraphics[width=40mm]{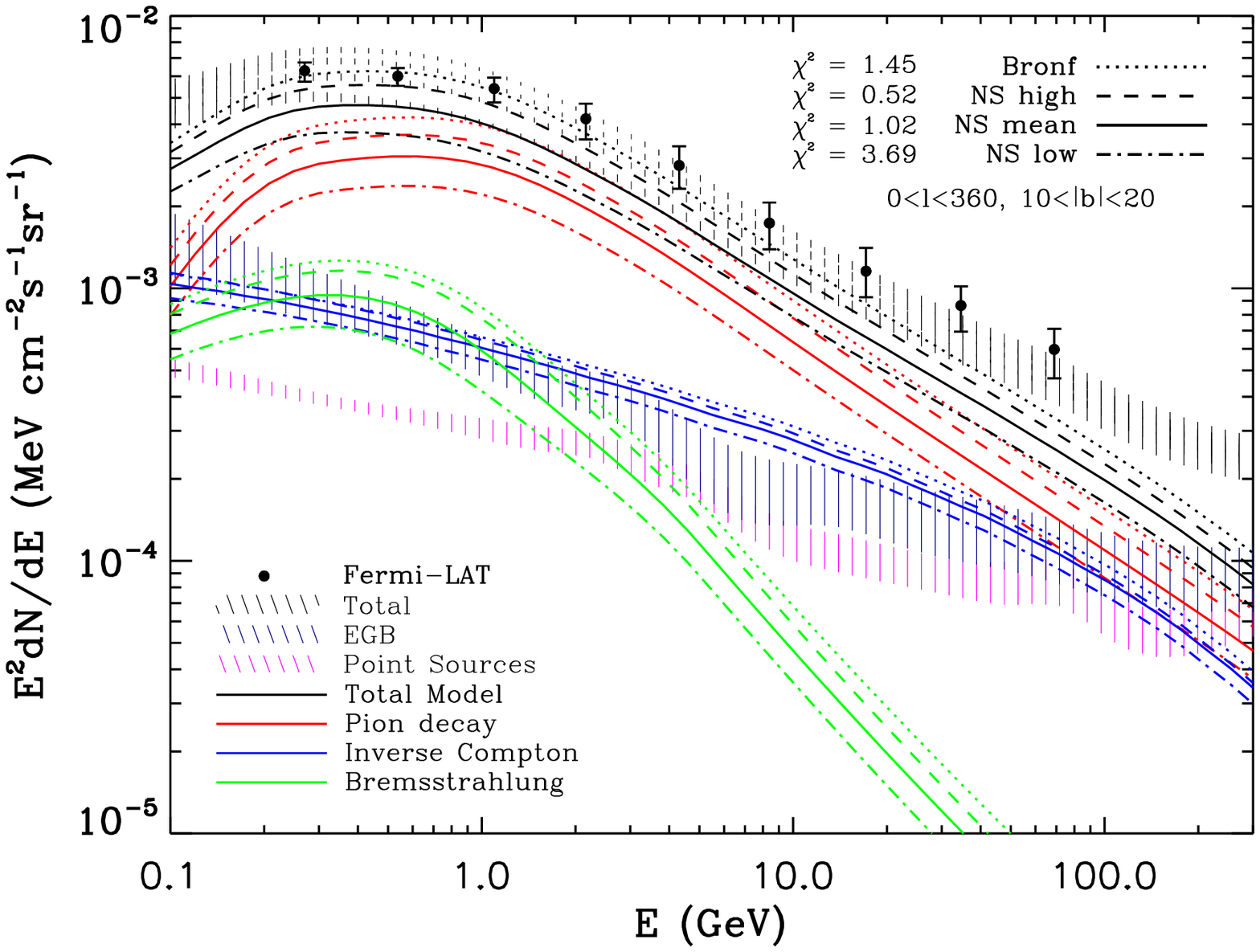}
\includegraphics[width=40mm]{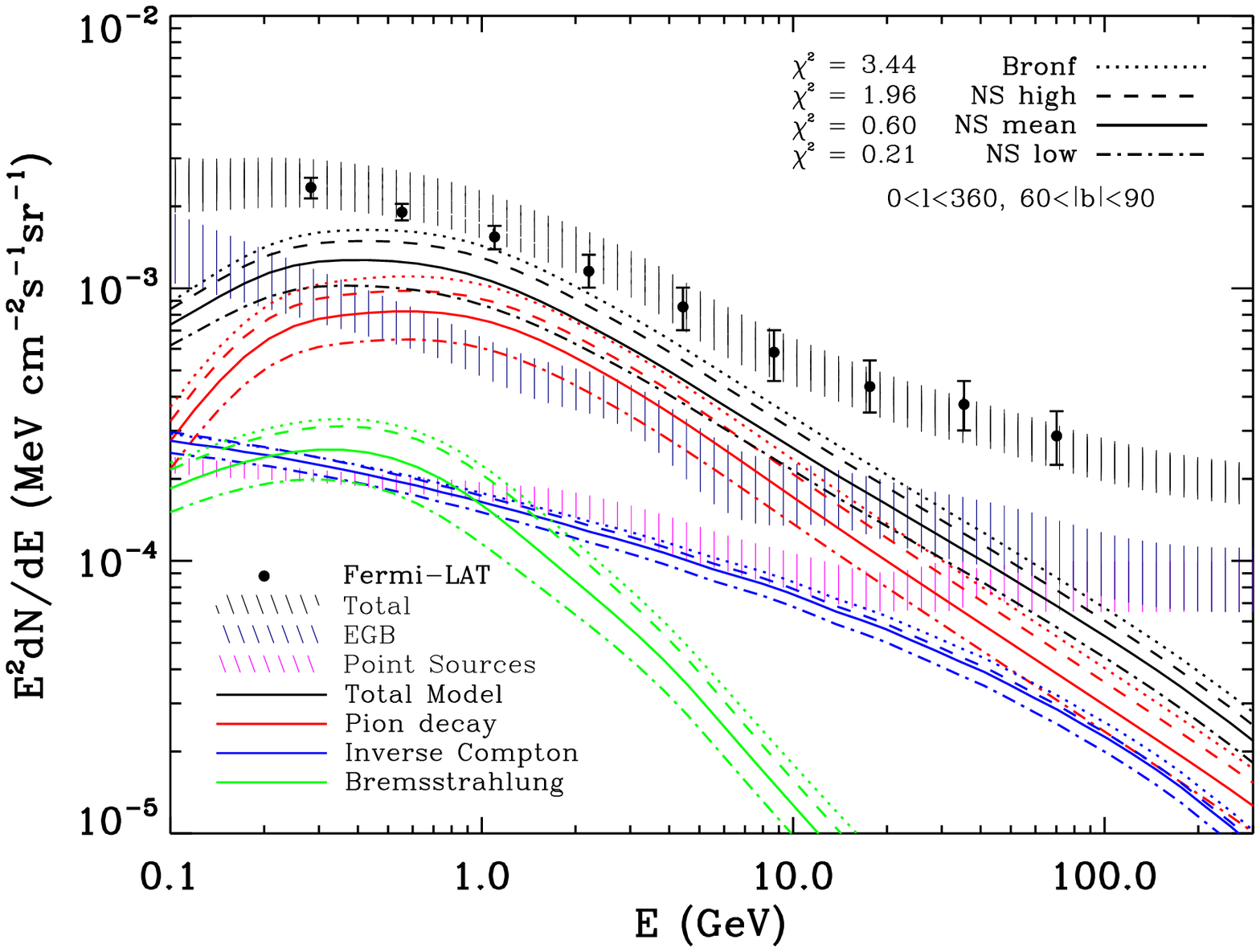}
\end{center}
\caption{Predictions for different molecular hydrogen distributions shown in Fig~\ref{fig:Gasprofile}.}
\label{fig:GasVary}
\end{figure}

\begin{table}
\begin{tabular}{|c|c|c|c|c|c|c|c|c|c|c|}
   \hline
   Name &$\delta$&$z_d$&$r_d$&$D_0$&$v_A$&$\eta$&$\gamma_1^p/\gamma_2^p/\gamma_3^p$&$R^p_1$\\
   \hline\hline
   \scriptsize{KRA4-20}&\scriptsize{0.5}&\scriptsize{4}&\scriptsize{20}&\scriptsize{2.49}&\scriptsize{19.5}&\scriptsize{-0.36}&\scriptsize{2.06/2.35/2.18}&\scriptsize{14.9}\\
   \hline
   \scriptsize{KRA4-5}&\scriptsize{0.5}&\scriptsize{4}&\scriptsize{5}&\scriptsize{2.76}&\scriptsize{16.9}&\scriptsize{0.0}&\scriptsize{2.07/2.35/2.18}&\scriptsize{27}\\
   \hline
   \scriptsize{KRA4-10}&\scriptsize{0.5}&\scriptsize{4}&\scriptsize{10}&\scriptsize{2.58}&\scriptsize{19.1}&\scriptsize{-0.25}&\scriptsize{2.05/2.35/2.18}&\scriptsize{17.5}\\
   \hline
   \scriptsize{KRA1-20}&\scriptsize{0.5}&\scriptsize{1}&\scriptsize{20}&\scriptsize{0.55}&\scriptsize{16.3}&\scriptsize{-0.52}&\scriptsize{2.07/2.34/2.18}&\scriptsize{16.5}\\
   \hline
   \scriptsize{KRA10-20}&\scriptsize{0.5}&\scriptsize{10}&\scriptsize{20}&\scriptsize{4.29}&\scriptsize{19.1}&\scriptsize{-0.37}&\scriptsize{2.05/2.35/2.18}&\scriptsize{15.2}\\
   \hline
   \scriptsize{RUN4-20}&\scriptsize{0.4}&\scriptsize{4}&\scriptsize{20}&\scriptsize{3.21}&\scriptsize{23.2}&\scriptsize{0.32}&\scriptsize{2.06/2.44/2.28}&\scriptsize{14}\\
   \hline   
   \scriptsize{KOL4-20}&\scriptsize{0.33}&\scriptsize{4}&\scriptsize{20}&\scriptsize{3.85}&\scriptsize{24.8}&\scriptsize{0.77}&\scriptsize{2.03/2.49/2.35}&\scriptsize{10.7}\\
   \hline
\end{tabular}
\caption{The parameters for the various models of propagation. See text, $z_{d}$ and $r_{d}$ in kpc, $D_{0}$ in $\times 10^{28} \textrm{cm}^2 \textrm{s}^{-1}$ (at 3 GV) and $v_{A}$ in km/s; $\gamma_1^p$ is the protons injection index below the $R_1^p$ (in GV), 
$\gamma_2^p$ the injection index between $R_1^p$ and $R_2^p = 300$~GV, and $\gamma_3^p$ above $R_2^p$. For primary electrons we assumed one break at 5~GV 
above(below) which, the injection index is 2.62(1.6). The pulsar parameters are $n=1.4$, $M=1.2~TeV$.}
\label{tab:Param}
\end{table}

Among different components of the ISG, molecular hydrogen distribution has large uncertainties (see Fig.~\ref{fig:Gasprofile}). Our reference for H2 distribution  \cite{Nakanishi:2006zf} is called "NS mean". 
To account for uncertainties in H2 midplane density we also study its high and low values called, respectively, "NS high" and "NS low". 
We study the model developed by  \cite{1988ApJ...324..248B} as well ("Bronf"). The larger number of target nuclei in the gas model \cite{1988ApJ...324..248B}, results in the need of a faster escape of CRs from the Galaxy, in order not to overproduce secondaries. 
As a result, to keep the same B/C flux ratio, the diffusion coefficient normalization needs to be increased. 
Therefore $e^{\pm}$ would propagate to larger distances from the Galactic disk, resulting in an  increase in the observed flux. Thus the combined analysis of CRs and $\gamma$-rays can constrain the large scale distribution of the gas in the Galaxy (see \cite{Cholis:2011un} for more details on ISM constraints). 

In conclusion, we have studied a rather extreme set of CR galactic diffusion profiles,
ranging from an equivalently constant galactic diffusion coefficient, to a diffusion 
profile with large gradients both in $r$  and $z$. 
We find that thicker diffusion zones are preferred by the combined fit of
CRs and $\gamma$-rays, while the $r$-dependence of the diffusion coefficient 
can not get strongly constrained from the current data. More importantly,
we find that even with the existing data, such a combined analysis can discriminate
and even constrain profiles of the ISM gas.  


\end{document}